\documentclass[prc,aps,eqsecnum,floatfix,showpacs]{revtex4}
\usepackage{dcolumn}
\usepackage{bm}
\usepackage{graphicx}
\begin{document}

\title{Induced Polarization in the $^2$H($\gamma,\vec n$)$^1$H Reaction at Low Energy} 
\author{R.\ Schiavilla}
\affiliation{Jefferson Lab, Newport News, Virginia 23606 \\
         Department of Physics, Old Dominion University, Norfolk, Virginia 23529}
\date{\today}

\begin{abstract}
The induced polarization, $P^\prime_y$, of the neutron in the deuteron photo-disintegration
from threshold up to 30 MeV is calculated using a variety of different, latest-generation
potentials---Argonne $v_{18}$, Bonn 2000, and Nijmegen I--- and a realistic model for
the nuclear electromagnetic current operator, including one- and two-body terms.  The
model dependence of the theoretical predictions is found to be very small.  These
predictions are systematically larger in magnitude than the measured $P^\prime_y$ values,
and corroborate the conclusions of an earlier, and much older, study.  There is considerable
scatter in the available experimental data.  New and more accurate measurements
of the induced polarization in the $^2$H($\gamma,\vec n$)$^1$H reaction are needed
in order to establish unequivocally whether there is a discrepancy between theory
and experiment.
\end{abstract}
\maketitle
\section{Introduction, Results, and Conclusions}
\label{sec:intro}

\begin{figure}[bthp]
\includegraphics[width=6in]{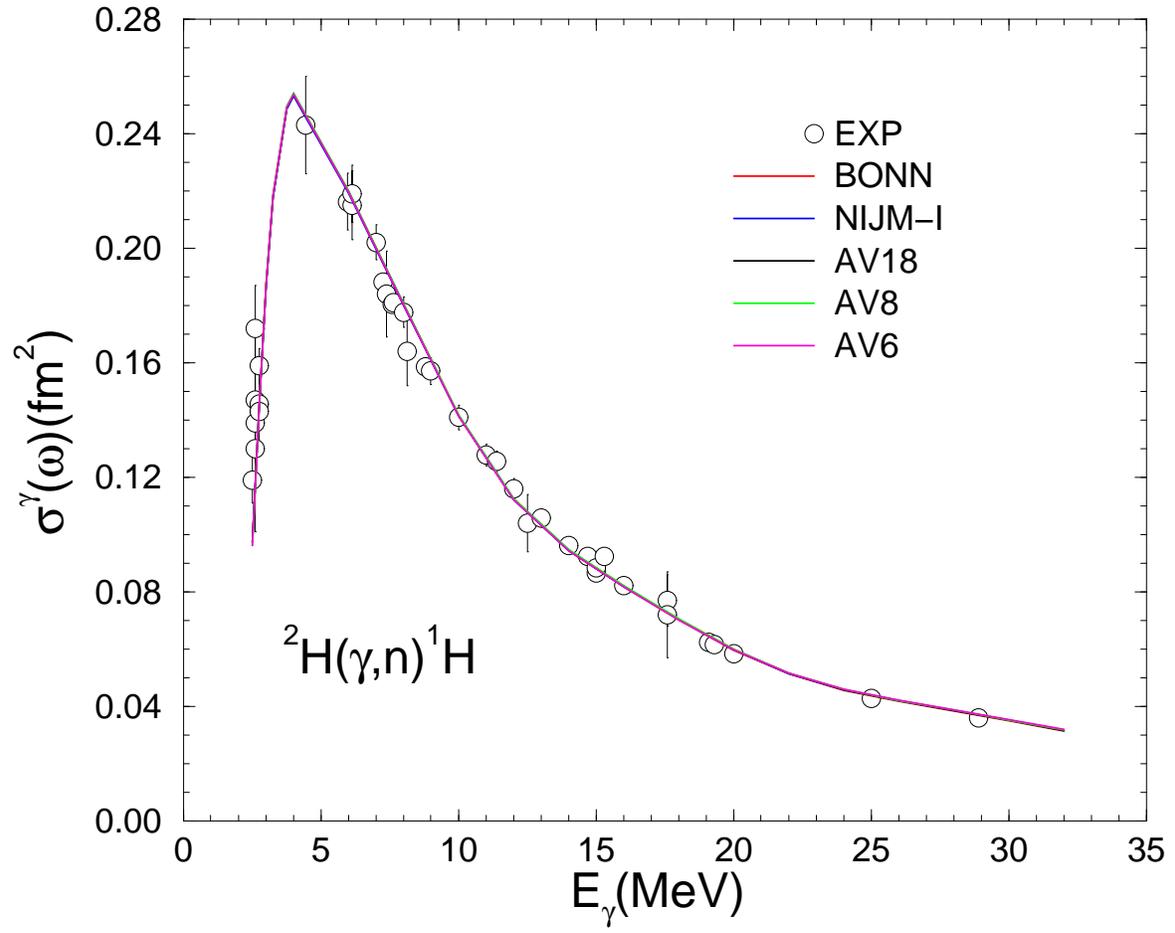}
\caption{(Color online) The deuteron photo-disintegration cross section, calculated
with a number of modern nucleon-nucleon potentials, is compared to data.  Note that
the various curves are indistinguishable.}
\label{fig:xs}
\end{figure}

\begin{figure}[bthp]
\includegraphics[width=6in]{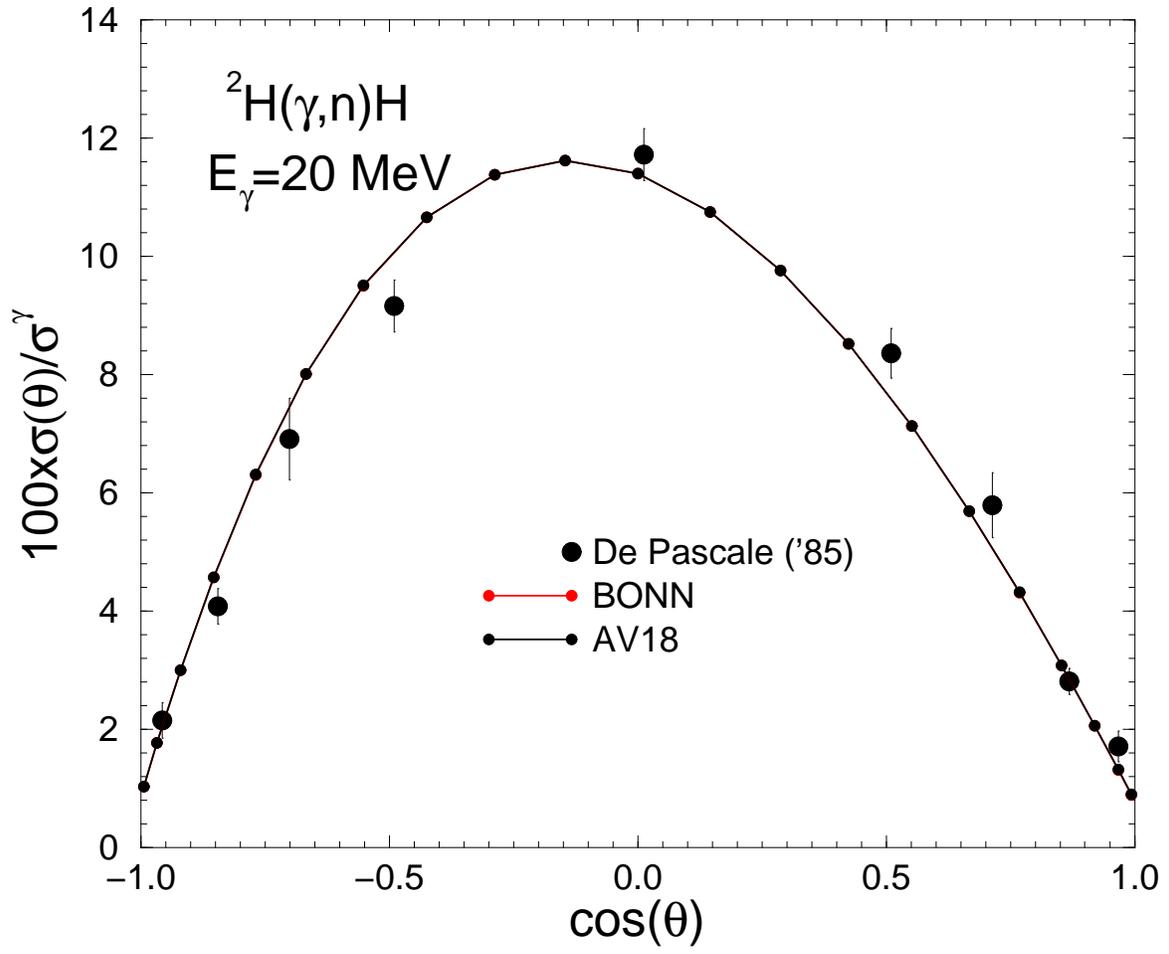}
\caption{(Color online) The center-of-mass angular distribution of the
experimental photo-disintegration cross section, normalized to the total
cross section, is compared to the results of calculations based on the BONN and AV18
potentials.  Note that the theoretical curves are indistinguishable.}
\label{fig:xs_20}
\end{figure}

\begin{figure}[bthp]
\includegraphics[width=6in]{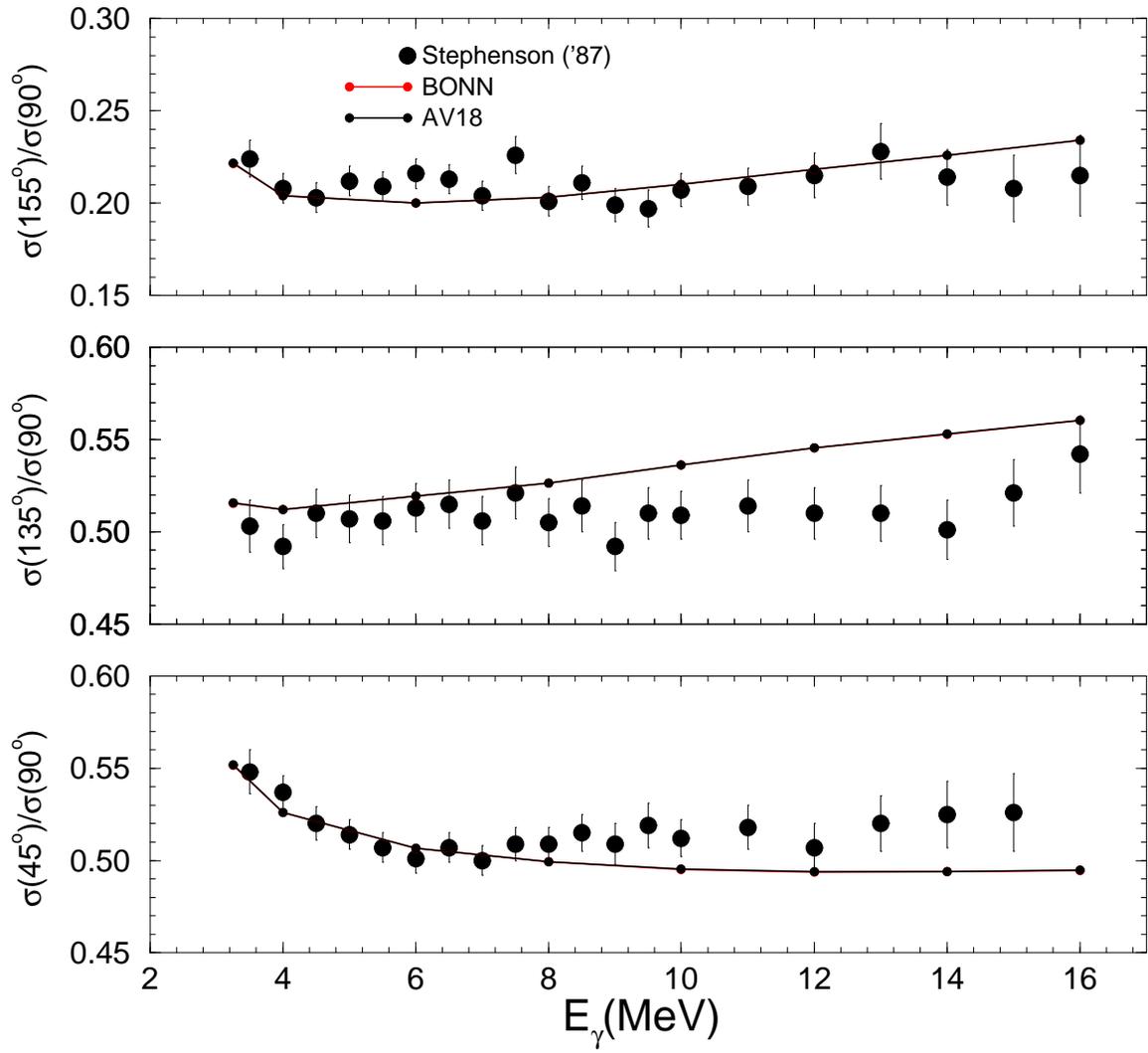}
\caption{(Color online) The lab-frame angular distribution ratios,
measured in the deuteron photo-disintegration as function of
photon energy, are compared to the results of calculations based on the BONN and AV18
potentials.  Note that the theoretical curves are indistinguishable.}
\label{fig:x_ratio}
\end{figure}

The present note deals with the deuteron photo-disintegration
and neutron induced polarization in the $^2$H($\gamma,\vec n$)$^1$H reaction at low
energy, from threshold up to about 30 MeV.  In this energy range, the photo-disintegration 
process is dominated by the contributions of electric dipole ($E1$) and, to a much less
but still significant extent, electric quadrupole ($E2$) transitions, connecting the deuteron
to the $n$$p$ $^3$P$_{J=0,1,2}$ and $^3$S$_1$--$^3$D$_1$ states, respectively.
The experimental data are well reproduced by theory, see Refs.~\cite{Schiavilla04,Marcucci05}
and Figs.~\ref{fig:xs}--\ref{fig:x_ratio}.  The total cross section data are from
Refs.~\cite{Bishop50,Snell50,Colgate51,Carver51,Ahrens74,Birebaum85,Bernabei86,Moreh89,DeGraeve92},
the angular distribution data at photon energy $E_\gamma$=19.8 MeV are from Ref.~\cite{DePascale85},
and the data on angular distribution ratios as function of $E_\gamma$ are from
Ref.~\cite{Stephenson87}.  The calculations shown in these figures, various aspects
of which are succintly summarized in the following section, are based
on a variety of (modern) realistic nucleon-nucleon potentials, including
the CD-Bonn (BONN)~\cite{Machleidt01}, Nijmegen I (NIJM-I)~\cite{Stoks94},
and Argonne $v_{18}$ (AV18)~\cite{Wiringa95}, as well as on semi-realistic
reductions of the AV18~\cite{Wiringa02}, the Argonne $v_6$ (AV6) and
Argonne $v_8$ (AV8) models, constrained to reproduce the binding energy
of the deuteron and the isoscalar combination of the S- and P-wave phase
shifts.  In particular, the AV6 ignores spin-orbit interaction components,
which are important in differentiating among the $^3$P$_{0,1,2}$ channels,
and therefore does not provide a good fit to the phase shifts in these channels.
The model dependence of all theoretical predictions shown in Figs.~\ref{fig:xs}--\ref{fig:x_ratio},
including those corresponding to the AV6 and AV8 in Fig.~\ref{fig:xs}, is negligible.

Calculations of the $n$$p$ radiative capture cross section at thermal
neutron energies, based on these same potential models, are also found
to be in excellent agreement with the measured value, when two-body
current contributions are taken into account~\cite{Schiavilla04,Marcucci05}.
The model dependence is again negligible.  The $n$$p$ radiative capture up
to neutron energies of about 100 keV proceeds almost exclusively through
the well-known magnetic dipole ($M1$) transition connecting the
$^1$S$_0$ $n$$p$ and deuteron states~\cite{Marcucci04}.

\begin{figure}[bthp]
\includegraphics[width=6in]{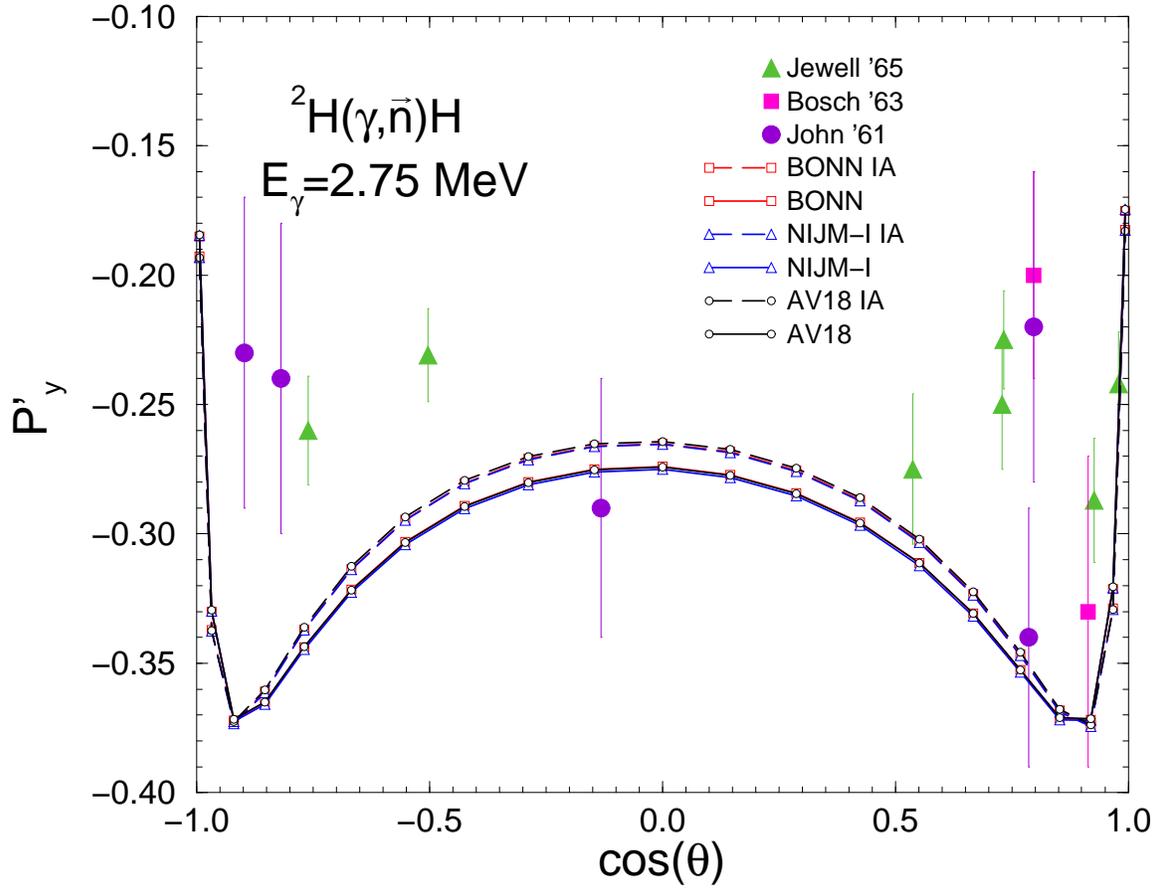}
\caption{(Color online) The center-of-mass angular distribution of the neutron
induced polarization measured in the $^2$H($\gamma,\vec n$)$^1$H
reaction at photon energies of 2.75 MeV
is compared to the results of calculations based on a number of latest-generation
nucleon-nucleon potentials and a realistic model for the nuclear electromagnetic
current, including one- and two-body components.  Also shown are the results
obtained by ignoring two-body currents (labeled IA).}
\label{fig:py_2.75}
\end{figure}

\begin{figure}[bthp]
\includegraphics[width=6in]{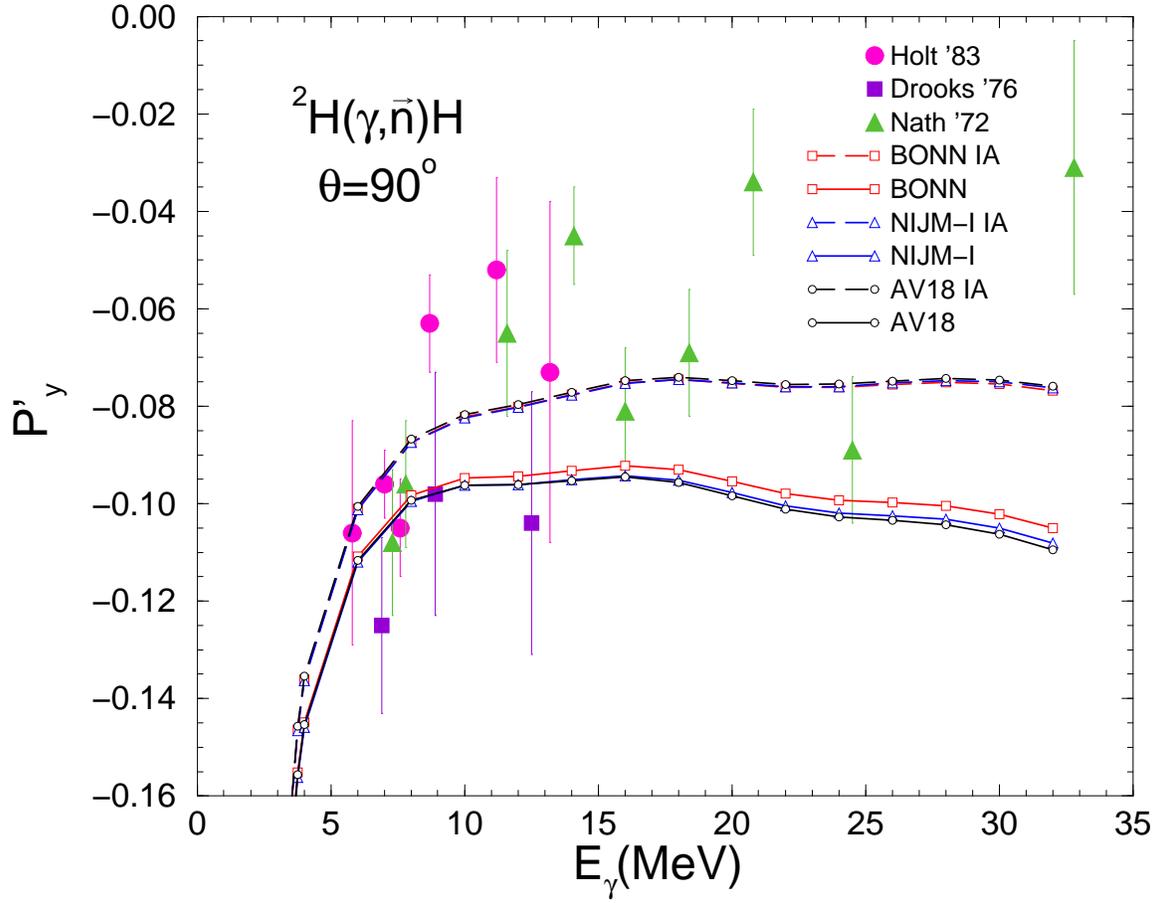}
\caption{(Color online) The neutron induced polarization measured in the
$^2$H($\gamma,\vec n$)$^1$H reaction at center-of-mass angle $\theta$=90$^\circ$ is
compared to the results of calculations based on a number of latest-generation
nucleon-nucleon potentials and a realistic model for the nuclear electromagnetic
current, including one- and two-body components.  Also shown are the results
obtained by ignoring two-body currents (labeled IA).}
\label{fig:py}
\end{figure}

\begin{figure}[bthp]
\includegraphics[width=6in]{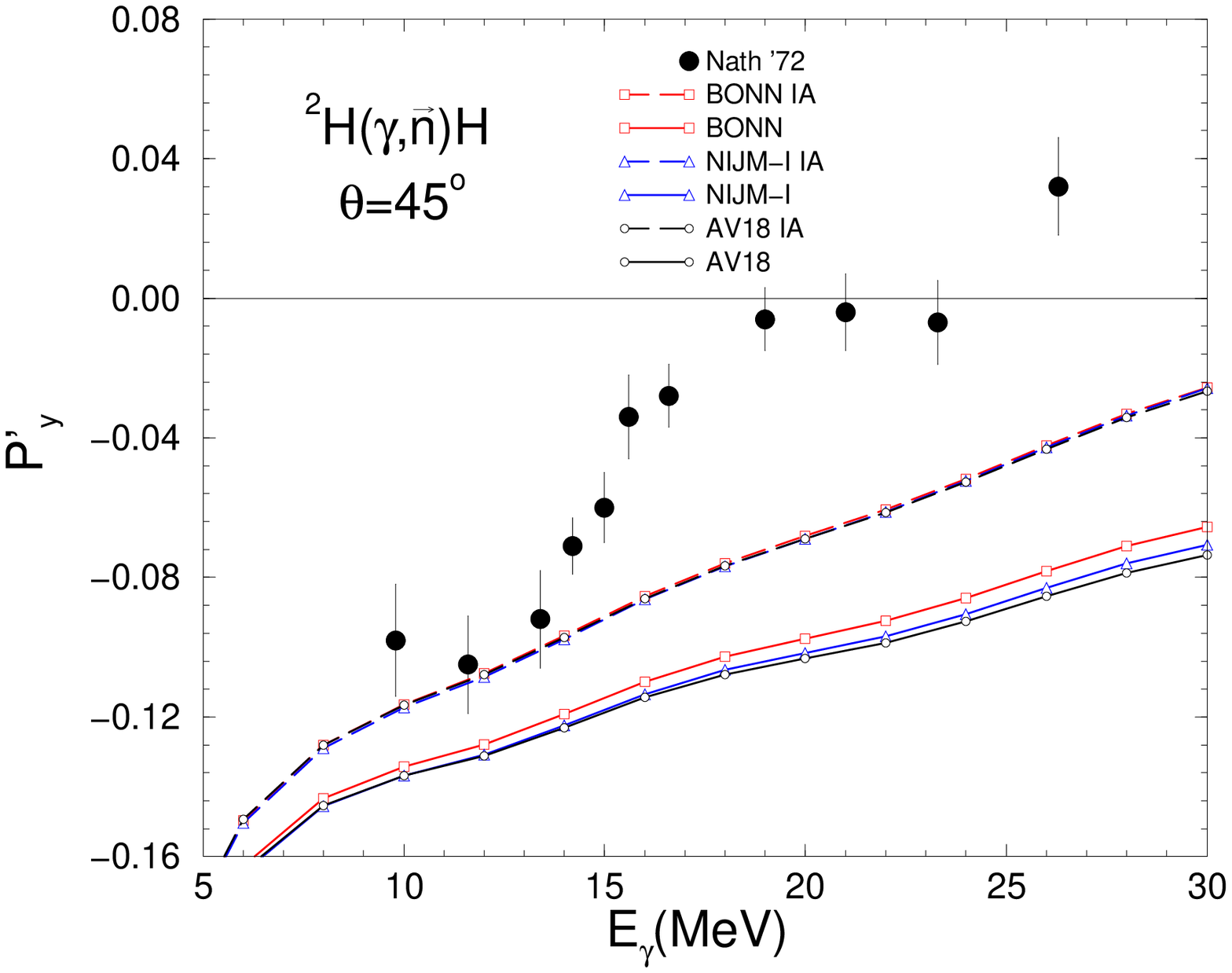}
\caption{(Color online) Same as in Fig.~\protect\ref{fig:py}, but at
center-of-mass angle $\theta$=45$^\circ$.}
\label{fig:py_45}
\end{figure}

\begin{figure}[bthp]
\includegraphics[width=6in]{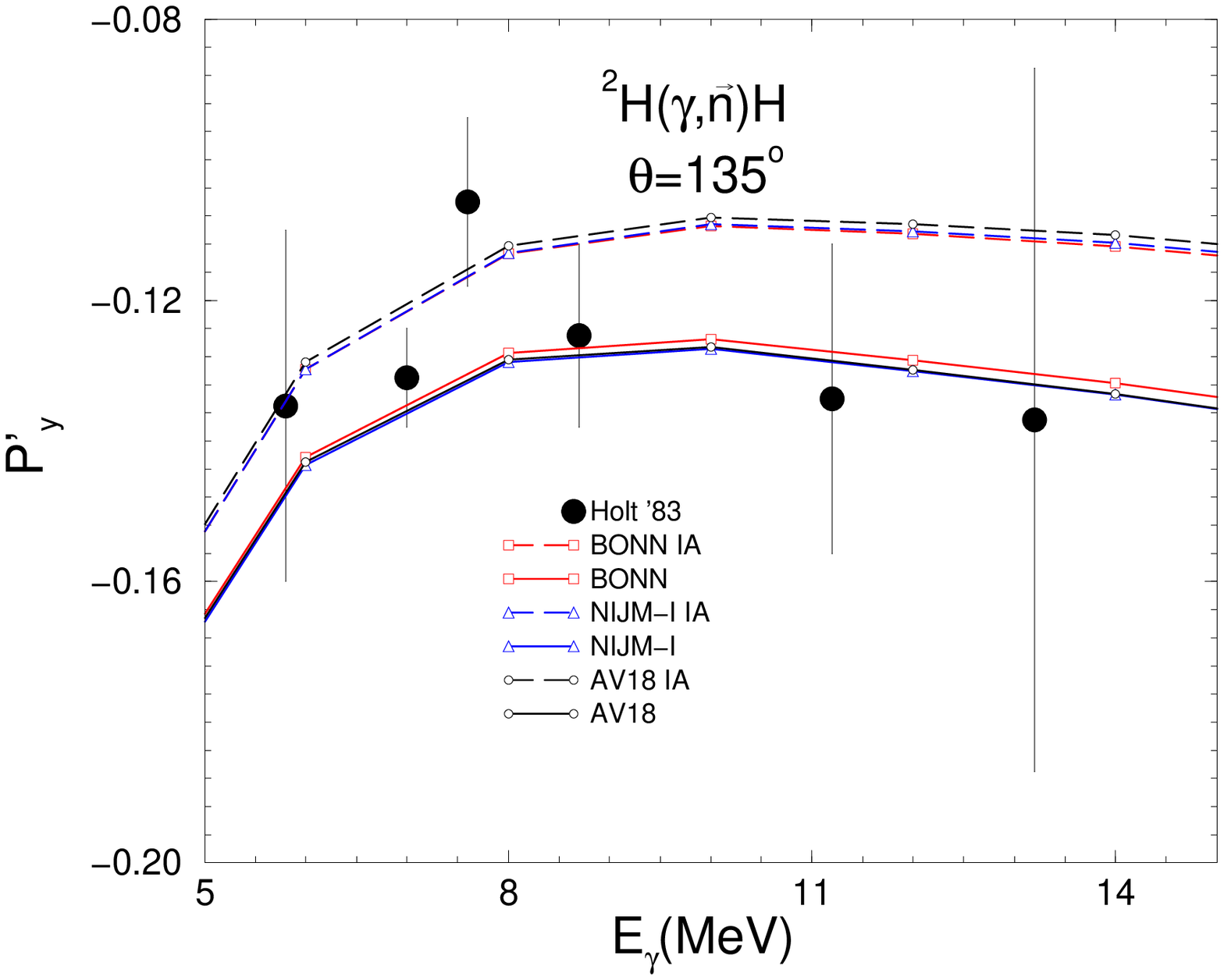}
\caption{(Color online) Same as in Fig.~\protect\ref{fig:py}, but at
center-of-mass angle $\theta$=135$^\circ$.}
\label{fig:py_135}
\end{figure}

On the basis of these facts, one is led to conclude that the $M1$ and $E1$
transition strengths, which the $n$$p$ radiative capture and deuteron
photo-disintegration are selectively sensitive to at low energies, are
both consistent with experimental data.  It is known~\cite{Rustgi60} that
the neutron induced polarization ($P^\prime_y$) in the $^2$H($\gamma,\vec n$)$^1$H
reaction originates predominantly, in the low-energy regime of interest here,
from interference of $M1$ and $E1$ transition terms.  Thus, it is surprising
to find that this observable, measured up to photon energies of 25 MeV,
is significantly overestimated, in magnitude, by theory, as shown in
Figs.~\ref{fig:py_2.75}--\ref{fig:py_45}, although the data at the (center-of-mass)
angle of 135$^\circ$ in Fig.~\ref{fig:py_135} seem to be consistent with it.  The
$P^\prime_y$ angular distribution data at $E_\gamma$=2.75 MeV
are from Refs.~\cite{John61,Bosch63,Jewell65}, while the data at $\theta$=90$^\circ$
for $E_\gamma$=6--30 MeV are from Refs.~\cite{Nath72,Drooks76,Holt83}, and
those at $\theta$=45$^\circ$ and 135$^\circ$ are from Refs.~\cite{Nath72}
and~\cite{Holt83,Holt05}, respectively.  In the figures the results obtained
without (IA) and with the inclusion of two-body current contributions are displayed
separately for the BONN, NIJM-I, and AV18 potential models.  The contributions of
two-body currents, essential if the observed cross section for the $n$$p$ radiative
capture is to be correctly predicted, turn out to substantially worsen the agreement
between the measured and calculated $P^\prime_y$ in all cases but at $\theta$=135$^\circ$.
The discrepancy between theory and experiment is particularly severe for
$P^\prime_y$ at $\theta$=45$^\circ$.

That the $P^\prime_y$ data are problematic for theory has in fact been known for
some time~\cite{Holt83,Holt05,Hadjimichael73,Schmitt91}.  Indeed, the main motivation
for the present study was to re-examine this issue in light of the advances
made during the last decade in the modeling of both nucleon-nucleon potentials
and two-body electromagnetic currents.  The corresponding results, however,
are close to those of Hadjimichael~\cite{Hadjimichael73}---reported in
Ref.~\cite{Holt83} at $\theta$=90$^\circ$---and Schmitt {\it et al.}~\cite{Schmitt91} and,
moreover, show a very small model dependence.  There is considerable scatter
among the different data sets at $\theta$=90$^\circ$ and in the measurements
of the $P^\prime_y$ angular distribution at $E_\gamma$=2.75 MeV.  Clearly,
more accurate data on both the energy dependence and angular distribution,
which could be used to isolate the multipole components, are needed in order
to resolve this confusing situation, and draw definite conclusions.  
\section{Calculation}
\label{sec:cal}
The relevant matrix element in the photo-disintegration of a deuteron in spin
projection $m_d$ initially at rest in the laboratory is

\begin{equation}
j^{(-)}_{m_n,m_p;\lambda,m_d}({\bf p},{\bf q}) =
^{(-)}\!\! \langle {\bf q};{\bf p},m_n,m_p \mid \hat{\bm \epsilon}_\lambda({\bf q}) \cdot
{\bf j}({\bf q})\mid m_d\rangle
\label{eq:jmin}
\end{equation}
where ${\bf q}$ is the momentum of the absorbed photon and
$\hat{\bm \epsilon}_\lambda$,$\lambda=\pm 1$, are the spherical
components of its polarization vector, ${\bf j}({\bf q})$ is the nuclear
electromagnetic current operator, and $\mid\! {\bf q};{\bf p},m_n,m_p\rangle^{(-)}$
represents an $n$$p$ scattering state with total momentum ${\bf q}$ and relative
momentum ${\bf p}$, satisfying incoming wave boundary conditions.  The $z$-axis
is taken along $\hat{\bf q}$, which also defines the spin-quantization axis.  In the
results of the calculations presented in Sec.~\ref{sec:intro}, the $n$$p$ state
includes all channels up to total angular momentum $J$=5, the contributions
of higher partial waves have been found numerically negligible.  The methods used to
solve for the two-nucleon bound- and scattering-state problems as well as the techniques
developed for the evaluation of the transition amplitudes above have been
described in considerable detail in Ref.~\cite{Schiavilla04}: they will
not be discussed further here.

It is convenient to introduce a second reference frame with axes ${\bf x}^\prime$,
${\bf y}^\prime$, and ${\bf z}^\prime$, in which the relative momentum
${\bf p}$ is along ${\bf z}^\prime$ with components
(${\rm sin}\theta\, {\rm cos}\phi,{\rm sin}\theta\, {\rm sin}\phi,{\rm cos}\theta$) 
with respect to the reference frame defined earlier.  The
${\bf x}^\prime$ and ${\bf y}^\prime$ axes are taken to have directions
(${\rm cos}\theta\, {\rm cos}\phi,{\rm cos}\theta\, {\rm sin}\phi,-{\rm sin}\theta$)
and ($-{\rm sin}\phi,{\rm cos}\phi,0$), respectively.  A neutron with
polarization in the $+{\bf y}^\prime$ direction, as an example, is represented by
the state 

\begin{equation}
\mid\!+{\bf y}^\prime\rangle = \frac{\mid +\rangle + {\rm i}\,
 {\rm e}^{ {\rm i}\phi}\mid -\rangle}{ \sqrt{2}} \ ,
\end{equation}
where $\mid \pm \rangle$ denote the spin states with $\pm 1/2$ projections
along $\hat{\bf z}$, i.e.  $\hat{\bf q}$.  The transition amplitude for emission
of a neutron with polarization in the $+{\bf y}^\prime$ direction is then obtained
from the linear combination

\begin{equation}
j^{(-)}_{+y^\prime,m_p;\lambda,m_d}=
\frac{1}{\sqrt{2}} \left[ j^{(-)}_{+, m_p;\lambda,m_d}
-{\rm i} {\rm e}^{-{\rm i}\phi} j^{(-)}_{-,m_p;\lambda,m_d} \right] \ .
\end{equation}
A similar expression holds for emission of a neutron with polarization
in the $-{\bf y}^\prime$ direction.  The induced polarization $P^\prime_y$ is
defined as

\begin{equation}
P^\prime_y=\frac{\sigma_{+y^\prime}(\theta)-\sigma_{-y^\prime}(\theta)}
         {\sigma_{+y^\prime}(\theta)+\sigma_{-y^\prime}(\theta)} \ ,
\end{equation}
where the differential cross section is given by

\begin{equation}
\sigma_{\pm y^\prime}(\theta) \equiv
\frac{ {\rm d}\sigma_{\pm y^\prime} } { {\rm d}\Omega } = 
\frac{\alpha}{24 \pi} \frac{m\, p}{q} \sum_{m_d,\lambda}
\sum_{m_p} \mid j^{(-)}_{\pm y^\prime,m_p;\lambda m_d}({\bf p},{\bf q}) \mid^2 \ .
\end{equation}
Here $\alpha$ is the fine structure constant, $m$ is the nucleon mass, and
the magnitude $p$ of the relative momentum is fixed by energy conservation.
The polarization parameters $P^\prime_x$ and $P^\prime_z$, proportional
to cross-section differences
for emission of neutrons with polarizations, respectively, in the $\pm \hat{\bf x}^\prime$
and $\pm \hat{\bf z}^\prime$ directions, vanish, as required by parity conservation,
and this fact has been explicitly verified in the numerical calculations.

The continuity equation allows one to express the nuclear electromagnetic
current entering Eq.~(\ref{eq:jmin}) as~\cite{Schiavilla04}

\begin{equation}
{\bf j}({\bf q})={\bf j}({\bf q})-{\bf j}({\bf q}\!=\!0)+{\rm i}\, 
\left[ H \, , \, \int{\rm d}{\bf x} \, {\bf x} \, \rho({\bf x}) \right] \ ,
\label{eq:j_s}
\end{equation}
where $H$ is nuclear Hamiltonian and $\rho({\bf x})$ is the nuclear charge
density operator.  This identity ensures that the Siegert form is used
for the $E1$ operator, dominant in the energy regime of interest here.  Note
that the matrix elements $j^{(-)}_{+y^\prime,m_p;\lambda,m_d}$ are calculated
as discussed in Ref.~\cite{Schiavilla04}, namely without performing the expansion of
${\bf j}({\bf q})$ in terms of electric and magnetic multipole operators.
Of course, the commutator term in Eq.~(\ref{eq:jmin}) reduces to

\begin{equation}
{\rm i} \int{\rm d}{\bf x} \, {\bf x}
\left[ H\, , \, \rho({\bf x}) \right] \rightarrow {\rm i} q \,
\int{\rm d}{\bf x} \, {\bf x} \, \rho({\bf x}) \equiv {\rm i}\, q \,{\bf d}\ ,
\end{equation}
when evaluating the matrix elements.
The dominant contribution ${\bf d}^{\rm NR}$ to the electric dipole operator
${\bf d}$, to which the (unretarded) $E1$ multipole operator
is proportional, is simply given by

\begin{equation}
{\bf d}^{\rm NR} = \sum_i P_i ({\bf r}_i -{\bf R}) \ ,
\end{equation} 
where ${\bf R} = ({\bf r}_1+{\bf r}_2)/2$ is the center-of-mass position vector
and $P_i$ is the proton projector.  However, there are a number of
relativistic corrections which have been included in the present study,
due to i) the spin-orbit term in the single-nucleon charge density operator,
ii) the leading two-body contribution to $\rho({\bf x})$, associated with pion
exchange, and iii) the center-of-energy correction.  Explicit expressions for
the associated operators can be found in Refs.~\cite{Viviani00,Nollett01}.
In particular, the center-of-energy correction arises because translationally
invariant wave functions require center-of-energy rather than center-of-mass
coordinates.  The correct electric dipole operator should be defined as

\begin{equation}
{\bf d} = \sum_i P_i ({\bf r}_i -{\bf R}_{\rm CE}) \ ,
\end{equation}
where

\begin{equation}
{\bf R}_{\rm CE}=\frac{1}{2} \left( \frac{1}{\sum_i E_i}
\sum_i E_i \, {\bf r}_i + {\rm h.c.} \right) \ ,
\end{equation}
and (for two particles) 

\begin{equation}
E_i\simeq m +\frac{{\bf p}_i^2}{2 m} + \frac{v_{12}}{2} \ .
\end{equation}
This leads to a correction term to ${\bf d}^{\rm NR}$
of the form

\begin{equation}  
{\bf d}^{\rm CE} = {\bf R} - {\bf R}_{\rm CE} \simeq -\frac{1}{8m^2}
\left[ \frac{1}{1+H/(2m)} ({\bf p}\cdot {\bf P}) {\bf r} + {\rm h.c.}\right] \ ,
\end{equation}
for transitions between states with zero $z$-component of the total isospin,
${\bf P}$ is the pair total momentum.
Quantitatively, however, this as well as the spin-orbit and pion-exchange
corrections to ${\bf d}^{\rm NR}$ have been found to be rather small
for $E_\gamma$ up to 30 MeV (see below).  It is interesting to note that
in the $^4$He($d,\gamma$)$^6$Li radiative capture, the matrix elements
of ${\bf d}^{\rm NR}$ vanish because of isospin selection rules (and the
use of translationally invariant wave functions), and the relativistic
corrections above are responsible for the $E1$ strength which dominates the
cross section for this process at energies of 100 keV and below~\cite{Nollett01}.

The two-body currents in the calculations based on the BONN and NIJM-I potentials 
include the terms associated with $\pi$- and $\rho$-meson exchanges, $\Delta$-excitation,
and $\rho$$\pi$$\gamma$ and $\omega$$\pi$$\gamma$ transition mechanisms (see
Ref.~\cite{Schiavilla04} and references therein).  The AV18
calculation includes, in addition, the two-body currents associated with its
momentum-dependent interaction components, as derived in Ref.~\cite{Marcucci05}.  It should
be emphasized that the AV18 currents are exactly conserved.

\begin{figure}[bthp]
\includegraphics[width=6in]{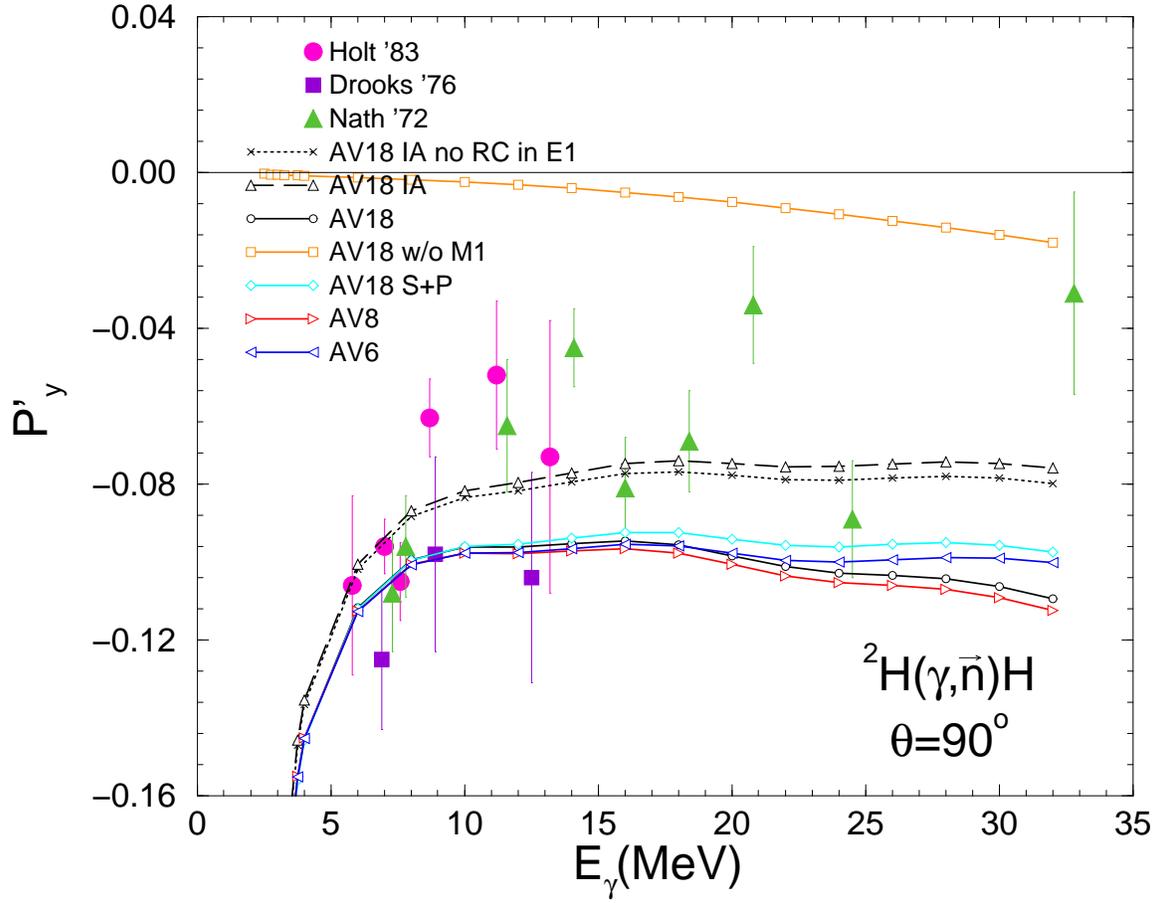}
\caption{(Color online) The neutron induced polarization measured in the
$^2$H($\gamma,\vec n$)$^1$H reaction at center-of-mass angle $\theta$=90$^\circ$ is
compared to results obtained in various approximation schemes (see text for
an explanation of the notation).}
\label{fig:py_cmp}
\end{figure}

Finally, Fig.~\ref{fig:py_cmp} is meant to illustrate the sensitivity of the
$P_y^\prime$ results to various inputs in the calculation.  The curves labeled
AV18, AV8 and AV6 denote the results of calculations based on the AV18, AV8
and AV6 potentials, respectively, including one- and two-body currents~\cite{Marcucci05} and the
relativistic corrections to the electric dipole operator discussed above (the
curves labeled BONN, NIJM-I and AV18 in Figs.~\ref{fig:xs}--\ref{fig:py_135} are obtained
in this same approximation scheme, while those labeled BONN IA, NIJM-I IA, and AV18 IA
ignore two-body current contributions).
As already emphasized, this polarization observable is not very sensitive to the
input potential, even in the case of a semi-realistic one like the AV6.  The
curves labeled AV18 IA and AV18 IA w/o RC in E1 both represent AV18-based
results including only one-body currents, the difference being that
in the AV18 IA w/o RC in E1 calculation the relativistic corrections to
${\bf d}_{\rm NR}$ are ignored, while in the AV18 IA one they are retained.
At the highest energy, the associated contributions reduce
(in magnitude) the AV18 IA results by about 5\% (in the total cross
section they decrease the AV18 IA values by less than 1\%
over the whole $E_\gamma$ range).
The curve labeled AV18 S+P shows the AV18-based results obtained by
including one- and two-body currents, but only S- and P-waves in the 
partial-wave expansion of the final $n$$p$ state.  As can be seen by
comparing the AV18 and AV18 S+P curves, the contributions of
the higher partial waves to $P_y^\prime$ are substantial at the highest photon energies.
Lastly, the results denoted as AV18 w/o M1 are obtained with the AV18, except
that the contribution of the $M1$ transition due to the $^1$S$_0$ state
has been neglected.  It demonstrates the sensitivity of the
$P_y^\prime$ observable at $\theta$=90$^\circ$ to this component of the
amplitude~\cite{Rustgi60}.  This sensitivity persists at
$\theta$=45$^\circ$ and 135$^\circ$.  Note that the total cross section
in Fig.~\ref{fig:xs} is reduced by less than 1\% for $E_\gamma$ in the 4-6 MeV range.
\section*{Acknowledgments}
The author wishes to thank Roy Holt for stimulating his interest in
the $^2$H($\gamma,\vec n$)$^1$H reaction, Ron Gilman and Roy Holt
for a critical reading of the manuscript, Gerry Hale and Ron Gilman
for making available to him experimental data sets of the deuteron
photo-disintegration cross section and neutron induced polarization
observable, respectively.  The author's work was supported by DOE
contract DE-AC05-84ER40150 under which the Southeastern Universities
Research Association (SURA) operates the Thomas Jefferson National
Accelerator Facility.  Finally, the calculations were made possible
by grants of computing time from the National Energy Research
Supercomputer Center.
%
%
%
 
%
%
%
%
\clearpage
\clearpage
\end{document}